\documentclass[showpacs,preprintnumbers,amsmath,amssymb,aps,twocolumn,floatfix]{revtex4}

\usepackage{bm}%
\usepackage{graphicx}
\usepackage{amsmath}
\usepackage{amssymb}
\usepackage{xspace}
\usepackage{color}

\newcommand{\InP}{InP:Fe\xspace}

\newcommand{\AffilMops}{\affiliation{Laboratoire Mat\'eriaux Optiques, Photonique et Syst\`emes \\ Unit\'e de Recherche Commune \`a l'Universit\'e Paul Verlaine -- Metz et Sup\'elec  \\ 2, rue Edouard Belin, 57070 Metz, France}}
\newcommand{\AffilPoma}{\affiliation{Laboratoire des Propri\'et\'es Optiques des Mat\'eriaux et Applications - CNRS UMR 6136\\ UFR Sciences, 2 Bd Lavoisier 49045 Angers, France}}

\newcommand{\ModifHerve}[1]{\textcolor{blue}{#1}}
\newcommand{\ModifNicolas}[1]{\textcolor{red}{#1}}
\renewcommand{\ModifHerve}[1]{#1}
\renewcommand{\ModifNicolas}[1]{#1}
\newcommand{\modif}[1]{\textcolor{red}{#1}}
\renewcommand{\modif}[1]{#1}
\newcommand{\modifh}[1]{\textcolor{blue}{#1}}
\renewcommand{\modifh}[1]{#1}

\begin{document}

\title{The theory of photorefractive resonance for localized beams \modif{in two-carrier photorefractive systems}}

\author{H.Leblond}
\AffilPoma

\author{N.Fressengeas}
\AffilMops


\begin{abstract}
    This paper extends the existing theory of two carrier photorefractivity \ModifNicolas{resonance}\modif{, which is generally applied to Iron doped Indium Phosphide (\InP),} to the case of low non-harmonic illumination.
\ModifHerve{The space charge field profile is computed, and the variations of its amplitude, width and position
are determined as functions of the background intensity. The effect of photorefractive resonance on these quantities is evidenced,}
 contributing to the understanding of published experimental results \modif{in \InP}.
\end{abstract}

\maketitle

\section{Introduction}
The photorefractive effect in iron doped Indium Phosphide (\InP) has been studied almost two decades ago using classical Two and Four Wave Mixing (TWM and FWM respectively) experiments\cite{Gla84apl44,mai88ol657}. These experiments have shown that \InP exhibits a photorefractive resonance at a given intensity, which enhances its TWM and FWM gain. This behavior has been successfully explained by a \modif{band transport two-carrier} theory\cite{pic89ol1362,Pic89ap3798} which is in full agreement with experiments: the resonance intensity $I_r$ is the intensity at which the thermal electron excitation rate is equal to the optically induced hole excitation rate.

TWM and FWM require a harmonic illumination, which allows to describe the induced photorefractive space charge field as a complex number: the real part is in phase with the illumination while the imaginary part is out of phase. In that case, the resonance is characterized by a maximum of imaginary part, corresponding to a maximum in the photorefractive TWM gain, and by a change of sign of the real part around $I_r$.

More recently, photorefractive self-focusing experiments in \InP have shown both a change of sign of the nonlinearity (from self-focusing to self-defocusing, or vice versa), along with a shift of the beam\cite{Cha96ol1333,cha97apl2499}. This observation has been naturally interpreted through a generalization of the resonance theory, where the local focusing effect is assumed to behave as the space charge field real part and where the beam shift in interpreted to reveal the imaginary part.

However, even more recent experiments\cite{fressengeas:063834}, using the same setup as in \cite{Cha96ol1333} but samples prepared in a different way, did not present the inversion of the self-focusing behavior or present it at intensity values different from the resonance intensity $I_r$ as measured from independent TWM experiments\cite{khel06oc169,fressengeas:063834}. \ModifNicolas{Those experiments were successfully described by a fully numerical model\cite{devaux:033823} which thus does not formally link the inversion to resonance.}

The goal of the present  paper is to formally generalize the theory of \cite{pic89ol1362,Pic89ap3798}, 
\ModifHerve{which we recall in section II,} to the case of a non-harmonic illumination such as the beam which is used in the self-focusing and spatial solitons experiments\cite{Cha96ol1333,cha97apl2499,fressengeas:063834}.

\modif{\section{Theory of photorefractive resonance for TWM in two-carrier photorefractive systems}}

\begin{figure}
 \includegraphics[width=\linewidth]{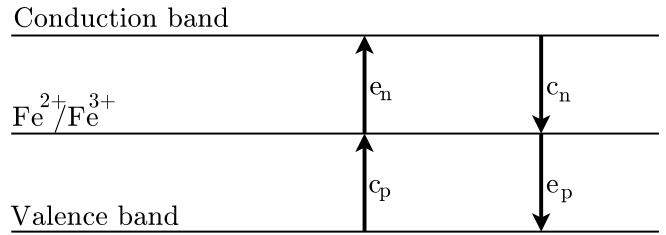}
 \caption{\modif{
Band transport two-carrier model with one middle band trap, characteristic, for instance, of Iron doped Indium Phosphide. The arrows shows the electrons motions between the energy levels with the associated coefficient as in  system (\ref{kuk}) and expressions (\ref{en}) and (\ref{ep}).}}
 \label{fig:InP}
\end{figure}

\modif{As gathered from the litterature\cite{Idr87oc317}, a band transport two-carrier model with a middle band trap can successfully describe photorefractivity in \InP, the middle band trap being the $\mathrm{Fe}^ {2+}/\mathrm{Fe}^ {3+}$ as shown on figure \ref{fig:InP}. If reduced to one dimension, this model is described by the following equation set:}

\begin{subequations}
\label{kuk}
\begin{align}
\frac {\partial E}{\partial x}&=\frac e\varepsilon\left(N_D-N_A+p-n-n_T\right),\label{kuk1}\\
j_n&=e\mu_n nE+\mu_nKT\frac{\partial n}{\partial x},\label{kuk2}\\
j_p&=e\mu_p pE-\mu_pKT\frac{\partial p}{\partial x},\label{kuk3}\\
\frac {\partial n}{\partial t}&=e_n n_T-c_n np_T+\frac1e\frac {\partial j_n}{\partial x},\label{kuk4}\\
\frac {\partial p}{\partial t}&=e_p p_T-c_p p n_T-\frac1e\frac {\partial j_p}{\partial x},\label{kuk5}\\
\frac {\partial n_T}{\partial t}&=e_p p_T-e_n n_T-c_p p n_T+c_n n p_T+\frac1e\frac {\partial j_n}{\partial x},\label{kuk6}\\
N_T&=n_T+p_T,\label{kuk7}
\end{align}
\end{subequations}

where $E$ is the electric field, $x$ and $t$ the space and time variables, $\varepsilon= \varepsilon_r\varepsilon_0$ the dielectric permittivity of
InP:Fe, $e$ the elementary electric charge, $N_D$, $N_A$, $n$, $p$ the densities  of donors, acceptors, free electrons, holes, respectively. $n_T$, $p_T$ and $N_T$ are the densities of iron in the form $\rm Fe^{2+}$, $\rm  Fe^{3+}$ and total, respectively.
$j_n$, $j_p$ are the current densities, and $\mu_n$, $\mu_p$ the mobilities, of electrons and holes, respectively. $K$ is Boltzmann's constant and $T$ the absolute temperature.
$e_n$, $ e_p$ are the excitation, $c_n$, $c_p$ the recombination coefficients of electrons and holes respectively.

The former write as
 \begin{eqnarray}
e_n=e_n^{th}+\sigma_n I,\label{en}\\
e_p=e_p^{th}+\sigma_p I,\label{ep}
\end{eqnarray}
where $e_n^{th}$, $e_p^{th}$ are the thermal, and $\sigma_n$, $\sigma_p$ the optical excitation  coefficients of electrons and holes respectively.

The theory of the photorefractive resonance is based on the assumption of a low fringe contrast \cite{pic89ol1362,Pic89ap3798}. In the case of single beam illumination, this hypothesis implies that the signal beam lies on a background intensity at least an order of magnitude more intense. This is however not the case in the previously published experiments\cite{Cha96ol1333,cha97apl2499,fressengeas:063834}  \ModifNicolas{nor in the recent numerical model\cite{devaux:033823}}. Furthermore, this uniform background beam \emph{cannot} account for any thermal generation, as was done previously for one carrier photorefractivity spatial solitons\cite{Fre96pre6866}.

However, as we will show, this hypothesis is \emph{absolutely} essential if one wants to generalize Picoli's resonance theory\cite{pic89ol1362,Pic89ap3798}. We will thus assume its validity throughout  this paper.


The intensity $I$ is written as
\begin{equation}
 I=I_0+I_1e^{ikx}+cc.,
\end{equation}
$I_1$ being written as $I_1=m I_0$, where $m$ is the fringe contrast. 
The low fringe contrast condition thus implies the assumption that $I_1\ll I_0$, or $m\ll 1$.

The photorefractive space charge electric field is expanded in the same way, as $E=E_0+E_1e^{ikx}+cc+\ldots$, where $E_0$ is the constant and uniform applied field,
 and the space charge field $E_1e^{ikx}$ the linear response to the small signal $I_1e^{ikx}$.
Performing the computation as in Refs. \cite{pic89ol1362,Pic89ap3798} but without any further approximation, we
\ModifHerve{retrieve the \ModifNicolas{same} expression \modif{as in refs \cite{pic89ol1362,Pic89ap3798}} for
the amplitude of the space charge field $E_1$}, \modif{which we recall here:}
\begin{widetext}
\begin{equation}
E_1=imI_0\frac{\sigma_pp_{T0}\left(1+\frac{i E_{p}}{E_0-i\left(E_d+E_{p}\right)}\right)
-\sigma_nn_{T0}\left(1-\frac{i E_{n}}{E_0+i\left(E_d+E_{n}\right)}\right)}
{e_{n0}n_{T0}\left(\frac1{E_q}+\frac{1-E_{n}/{E_q}}{\left(E_d+E_{n}\right)-iE_0}\right)+
e_{p0}p_{T0}\left(\frac1{E_q}+\frac{1-E_{p}/{E_q}}{\left(E_d+E_{p}\right)+iE_0}\right)}\label{eq_e}
\end{equation}
\end{widetext}
where
\begin{subequations}
\label{eqE}
\begin{align}
E_d&=k\frac{KT}e,\label{eqEd}\\
E_q&=\frac e{\varepsilon k}\frac{n_{T0}p_{T0}}{n_{T0}+p_{T0}},\label{eqEq}\\
E_n&=\frac{c_np_{T0}}{\mu_n k},\label{eqEn}\\
E_p&=\frac{c_p n_{T0}}{\mu_p k},\label{eqEp}
\end{align}
\end{subequations}
 $e_{n0}$, $e_{p0}$ are given by Eqs. (\ref{en},\ref{ep}) with $I=I_0$, and $n_{T0}$ and $p_{T0}$ are the value of the corresponding densities in the dark.

A first remark is that, as pointed out in the original paper \cite{Pic89ap3798}, Eq. (\ref{eq_e}) can be reduced to the \modif{same} 
form \modif{as in \cite{Pic89ap3798} :}
\begin{equation}
E_1=\frac{imI_0}
{\left(I_r+I_0\right)\left(\frac 1{E_q}+\frac{E_d}{E_0^2+E_d^2}\right)
+i\left(I_r-I_0\right)\frac{E_0}{E_0^2+E_d^2}}\label{eq_ered}
\end{equation}
which  evidences the resonance intensity
\begin{equation}
I_r=\frac{e_n^{th}n_{T0}}{\sigma_pp_{T0}},\label{eq_ires}
\end{equation}
\underline{only if} the mobility fields $E_n$ end $E_p$ are negligible with respect to the diffusion field $E_d$.

Notice that the space charge field $E_1$ is proportional to  the modulation $I_1=mI_0$, while the
resonance condition involves the background intensity $I_0$ only. The apparent relation between the amplitude of the modulation
$I_1$
and the resonance condition $I_0=I_{r}$ is an artifact due to the introduction of the fringe contrast $m=I_1/I_0$.
Erroneous interpretations may arise if one forgets that the intensity of the modulation is measured by $m$, and not by $I_0$, and that $m$ is assumed to be small.
Numerical investigation of the response to strong modulations, with a fringe contrast $m$ close to 1,
can be found in Ref. \cite{wolffer:6375}, but nowhere else to the best of our knowledge.


\section{Theory of photorefractive resonance for a localized beam}
We consider an illumination pattern $I_1(x)$ which is not harmonic any more, and assume that it is small if compared to a uniform background illumination $I_0$, as in the case of the TWM~\cite{pic89ol1362,Pic89ap3798}. The Kuhktarev equations (\ref{kuk}) can thus be linearized about the solution corresponding to the uniform illumination $I_0$, and
 solved in the general case by means of a Fourier transform, according to
\begin{equation}
E_1(x)=\int\hat E_1(k)e^{i k x} dk,
\end{equation}
 $\hat E_1(k)$ being given by formulas (\ref{eq_e},\ref{eqE}) above, which reduce to (\ref{eq_ered},
\ref{eq_ires}) when mobilities are assumed to be large and diffusion weak.
To let the $k$ dependency appear explicitly, let us simplify equations (\ref{eqEd},\ref{eqEq}) by setting
$E_q=E_0/ k r_1$ and $E_d= E_0 k r_2$,
with
\begin{equation}
r_1=\frac{\varepsilon E_0\left(n_{T0}+p_{T0}\right)}{e n_{T0}p_{T0}},
\end{equation}
and
\begin{equation}
r_2=\frac{K T}{eE_0}.
\end{equation}
Assuming $E_0=10 \,\rm kVcm^{-1}$, $T=297\,\rm K$, $n_{T0}=5\cdot 10^{15}\,\rm cm^{-3}$, $p_{T0}=6\cdot 10^{16}\,\rm cm^{-3}$,
and $\varepsilon_r=12.6$, which are reasonable values for InP:Fe\cite{Val88ap233,pic89ol1362}, we get
 $r_1=0.15\,\rm \mu  m$ and $r_2=0.026 \,\rm \mu m$.

Following, we \ModifHerve{obtain} from (\ref{eq_ered})  the following expression for   $\hat E_1(k)$ and any $I_0$~:
\begin{equation}
\hat E_1=\frac{iE_0\hat I_1}{\left(I_{r}+I_0\right)\left(k r_1+\frac{k r_2}{1+ k^2 r_2^2}\right)
+i\left(I_{r}-I_0\right)\frac1{1+ k^2 r_2^2}}.\label{eqE1_gen}
\end{equation}
Approximate expressions can \ModifNicolas{thus} be given for a background intensity $I_0$ well above, well below and precisely at resonance,  as the following:
\begin{subequations}
\begin{align}
\hat E_1&=\frac{iE_0}{I_0}\frac{\hat I_1}{\left(k r_1-\frac i{1-i k r_2}\right)}, \,&\mathrm{for}\, I_0\gg I_r,\\
\hat E_1&=\frac{iE_0}{I_r}\frac{\hat I_1}{\left(k r_1+\frac i{1+i k r_2}\right)}, \,&\mathrm{for}\, I_0\ll I_r,\,\\
\hat E_1&=\frac{iE_0}{2I_0}\frac{\hat I_1}{k r_1+\frac{k r_2}{1+ k^2 r_2^2}},\,&\mathrm{for}\, I_0=I_r.
\end{align}
\end{subequations}

Notice that the latter expression is singular for $k=0$.

Then, for any given $I_1(x)$, we compute its Fourier transform, report it into the above
 expressions, and compute the inverse Fourier transform $E_1(x)$.
In general, the computation can be performed numerically by means of a standard fast Fourier transform (FFT)
algorithm.
If $I_1(x)$ has a Gaussian shape, say \ModifNicolas{for instance}
\modifh{
\begin{equation}
I_1=Ae^{\frac{-x^2}{r^2}},
\end{equation}}
the inverse Fourier transform can be evaluated by approximately computing the integral by means of the
saddle point method, \emph{i.e.}, for $r$ large enough, by replacing the response
function $\hat E_1/\hat I_1$ by an expansion in a series of powers of $k$.
\begin{widetext}
We get, for $I_0\gg I_r$, up to the second order in $k$~:
\begin{equation}
E_1 = \frac{E_0A}{I_0}e^{-\frac{x^2}{r^2}}\left(-1 +
     \frac{2x}{r^2}(r_1 +r_2)
+\frac{2\left(r^2 - 2x^2\right)r_1(r_1 +2r_2)}{r^4}
     \right),\label{e1apup}
     \end{equation}
\ModifNicolas{and \ModifHerve{for $I_0\ll I_r$} in the same conditions:}
\begin{equation}
E_1 = \frac{E_0A}{I_r}e^{-\frac{x^2}{r^2}}\left(
1 + \frac{2x}{r^2}(r_1 +r_2)
    -\frac{2\left(r^2 - 2x^2\right)r_1(r_1 +2r_2)}{r^4}
     \right).\label{e1apdo}
\end{equation}
\end{widetext}

\begin{figure}
\begin{center}
\includegraphics[width=7cm]{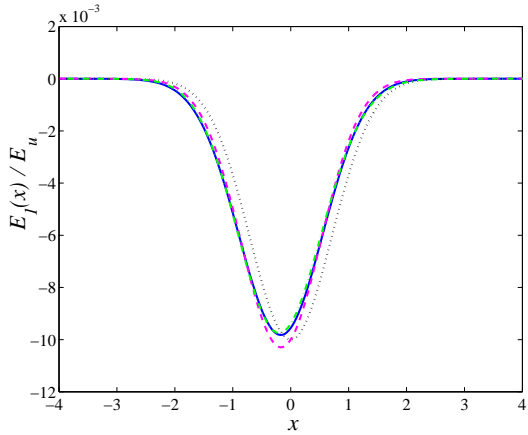}
\caption{\label{f1}
(Color online)
The profile of the \modifh{normalized space charge field $E_1(x)/E_u$}, for $I_0/I_{r}=100$, and $r= 1 \mu\rm m$.
Solid blue~: numerical, dashed pink: first saddle point method, dash-dotted green:second order one.
Dotted black: the intensity signal $-I_1$ \modifh{(divided by $AI_0/I_r$)}. Numerical box width: $2x_m=40$, number of points: 1024.
$x$ in $\mu$m. 
 }
\end{center}
\end{figure}

\begin{figure}
\begin{center}
\includegraphics[width=7cm]{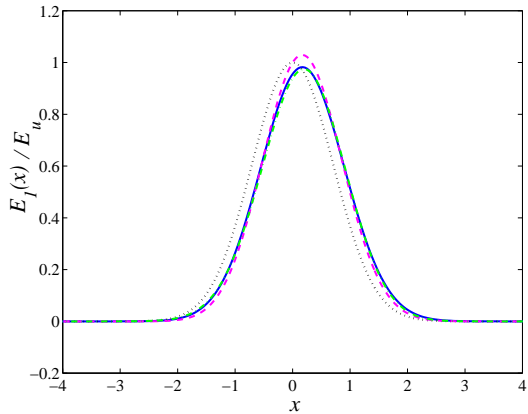}
\caption{\label{f2}
(Color online)
Same as fig. \ref{f1}, but $I_0/I_{r}=0.01 $. Dotted black line is \modifh{$+I_1/A$.}
$x$ in $\mu$m. 
 }
\end{center}
\end{figure}

Using the values of  $r_1$ and $r_2$ mentioned above, the first order formulas (\emph{i.e.} Eqs. (\ref{e1apup},\ref{e1apdo})
in which the term proportional to $1/r^4$ is neglected) are in good agreement with the exact numerical
solution down for  $r\gtrsim3\mu\rm m$, and for $r\gtrsim 1\mu\rm m$ if second order formulas are used,
see Figs. \ref{f1}-\ref{f2}. \modifh{ In these figures, as in the following ones, the space charge field $E_1$ is
normalized  by $E_u=E_0A/I_r$.
The figures are plotted for $r=1\mu\rm m$.For larger values of $r$, the figures have the same shape, but the
curves corresponding to the various computation methods are close together,
showing the higher accuracy of  the approximations.
}

Close to the resonance, the singularity in $1/k$ raises a difficulty, which can be solved using
the fact that
   \begin{equation}
   \frac{\partial E_1}{\partial x}={\cal F}^{-1}\left(-ik \hat E_1\right),
   \end{equation}
which yields, through the saddle point method,
\begin{equation}
   E_1 =  \frac{E_0A}{I_0}\left(   \frac{r_2^3 x}{\left(r_1 + r_2\right)^2r^2} e^{-\frac{x^2}{r^2}}
 +\frac{\sqrt\pi\; r}{4(r_1 + r_2)}{\rm erf}\left(\frac x r\right)\right),\label{app_reso}
 \end{equation}
where erf denotes the error function.

\begin{figure}
\begin{center}
\includegraphics[width=7cm]{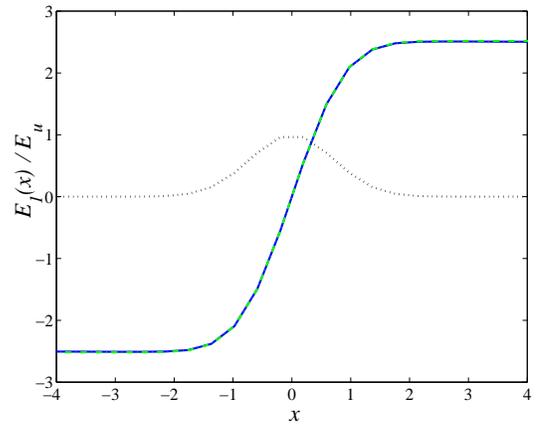}
\caption{\label{f3}
(Color online)
The profile of the \modifh{normalized space charge field $E_1(x)/E_u$}  for  $I_0=I_r$, and $r= 1 \mu\rm m$. Solid blue: numerical,
dash-dotted green: second order saddle point, dotted black: \modifh{$I_1/A$}.
Numerical box width: $2x_m=1600,$ number of points:  4096.
$x$ in $\mu$m. 
}
\end{center}
\end{figure}

Compared with the numerical solution  (in which $\hat E_1 (k=0)$ is set to zero to avoid  singularity),
the agreement is reasonable, except that the constant term is lost, and a linear term, which
depends on the box size, and hence can be considered as a numerical artifact.
The agreement holds down to  $r\simeq 0.1 \,\rm \mu m$ as can be seen on Fig. \ref{f3}.

\begin{figure}
\begin{center}
\includegraphics[width=7cm]{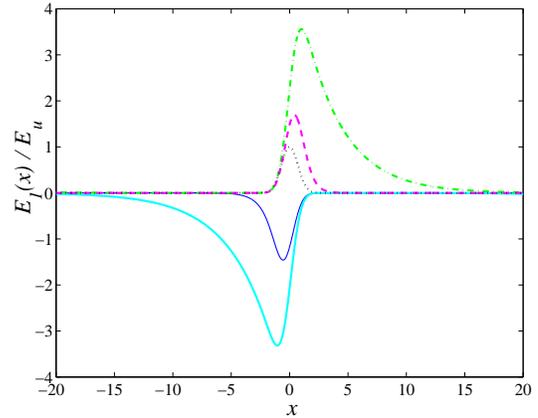}
\caption{\label{f4}
(Color online)
The profile of the \modifh{normalized space charge field $E_1(x)/E_u$}. Thin solid dark blue: $I_0/I_r=1.5$,
thick solid light  blue: 1.1, dash-dotted green: 0.9, dashed pink: 0.5.
Dotted black line is \modifh{$I_1/A$}.
$r= 1 \mu\rm m$. Numerical, with
 box width: $2x_m=40$, number of points:  1024. $x$ in $\mu$m.
}
\end{center}
\end{figure}

For other values of the ratio $I_0/I_r$ (\emph{i.e.} close to 1), no approximate analytic expression can be given, but the space charge field $E_1$
can be computed numerically using equation (\ref{eqE1_gen}) and a Fast Fourier Transform algorithm. A few examples are shown on Fig. \ref{f4}, evidencing the nature of the transition from a positive response of the field to a negative one when \ModifNicolas{the background illumination crosses the resonance intensity}.
\modifh{For $r=10\mu\rm m$, the same curve shape is obtained for ratios $I_0/I_r=1.05$, 1.01, 0.99, and $0.95$.}


\ModifHerve
{
>From the numerical computation of the space charge field $E_1$,
the variations of some characteristics of the latter can be computed, and drawn against the background intensity $I_0$. \ModifNicolas{We chose to  compute the evolution of those characteristics which are relevant for understanding the behavior of photorefractive self-focusing and spatial solitons: maximum space charge field value, its width with respect to the beam and its spatial displacement.}}

\ModifNicolas{
Figure \ref{figur} shows the  maximum absolute value $E_{1 max}$ of $E_1$  vs $I_0$.}
\ModifNicolas{As expected, i}\ModifHerve{t is a typical resonance curve, showing the photorefractive resonance at $I_0=I_r$.
\begin{figure}
\begin{center}
\includegraphics[width=7cm]{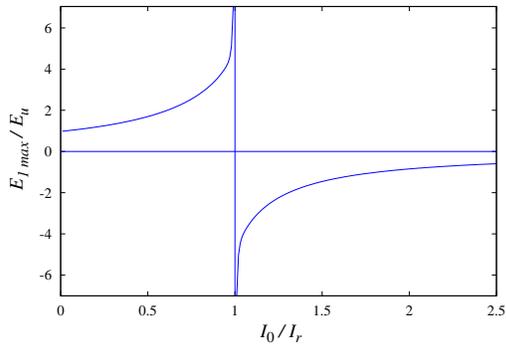}
\caption{\label{figur}
The extremal value \modifh{$E_{1 max}/E_u$ of the normalized} space charge field vs the normalized  background intensity $I_0/I_r$.
The resonance at $I_0=I_r$ clearly appears. 
}
\end{center}
\end{figure}
}

\ModifHerve
{
The effect of the resonance on the spatial extension of the space charge field is illustrated on Fig. \ref{larg}.
The width $L$ of the space charge field is computed as 
\begin{equation}
L=\sqrt{\langle x^2\rangle-\langle x\rangle^2},
\end{equation}
 with
\begin{equation}
\langle x^j\rangle=\frac{\int_{-\infty}^\infty x^j \left|E_1(x)\right|^2dx}{\int_{-\infty}^\infty  x^j \left|E_1(x)\right|^2dx}.
\end{equation}
At resonance, $I_0=I_r$, the width $L$ diverges, in accordance with expression (\ref{app_reso}) and Fig. \ref{f3}.
\begin{figure}
\begin{center}
\includegraphics[width=7cm]{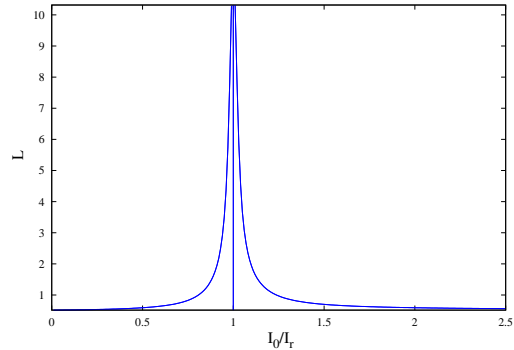}
\caption{\label{larg}
The width $L$ \modifh{(in $\mu$m)} of the space charge field vs the normalized  background intensity $I_0/I_r$.
The resonance at $I_0=I_r$ clearly corresponds to a critical widening of the screening.
}
\end{center}
\end{figure}
}

\ModifHerve{
The \ModifNicolas{displacement of the space charge field} with respect to the center of the beam is illustrated on Fig. \ref{dec}. The positions $\Delta x_+$ of the maximum of $E_1$, and $\Delta x_-$ of its minimum
are plotted vs $I_0$.
For $I_0<I_r$, the space charge field $E_1$ is positive, and $\Delta x_+$ is the location of its maximum,
it is positive: the space charge field is shifted in the direction of the external field.
If on the contrary $I_0<I_r$, then $E_1<0$ and its location is determined by $\Delta x_-$,
which is negative: the space charge field is shifted in the opposite direction. 
For  $I_0\ModifNicolas{=}I_r$, the location $\Delta x_+$ of the maximum of $E_1$ is further shifted to the right,
but it does not represent the maximal value of $|E_1|$ any more, and hence we did not report it
 on Fig. \ref{dec}.
\begin{figure} 
\begin{center}
\includegraphics[width=7cm]{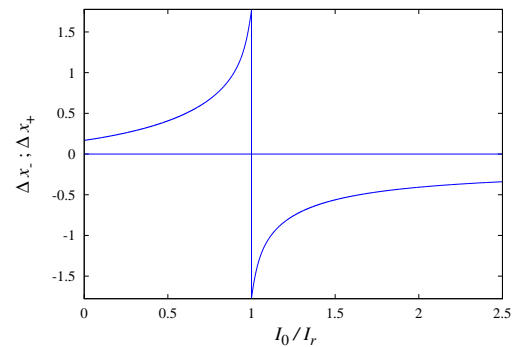}
\caption{\label{dec}
The positions $\Delta x_+$ (upper curve) and $\Delta x_-$ (lower curve) of the maximum and minimum of the space charge field  $E_1$ vs the normalized  background intensity $I_0/I_r$. The center of the beam is located at $x=0$.\modifh{  ($\Delta x_\pm$ in $\mu$m).}
}
\end{center}
\end{figure}
}
\modifh{For higher $r$, the curves on Figs. \ref{figur}, \ref{larg} and \ref{dec} have the same shape, except that they go closer to the asymptote.}

\section{Conclusion}

We have generalized to the general case and particularly to the case of a localized beam the theory of the photorefractive resonance in \modif{two-carrier photorefractive systems such as} Iron doped Indium Phosphide\cite{pic89ol1362,Pic89ap3798}, which
pertains only to two-wave and four-wave mixing experiments and the harmonic illumination they produce.

As could be inferred from the the harmonic theory\cite{pic89ol1362,Pic89ap3798}, the photorefractive space charge field indeed changes its sign around resonance.
\ModifHerve{
The amplitude, width and location of the space charge field 
show a singularity and/or an inversion of their behavior \ModifNicolas{around} the resonance.}
However, \ModifNicolas{and in contrast to previously thought,} the resonance condition does not compare the resonance intensity to the intensity of the localized signal itself, but to the intensity of a uniform background illumination, which is itself not so far from the mean intensity of the overall illumination of the sample.

Furthermore, this background intensity is assumed to be large with respect to the signal one.
This shows that, in the experiments in which only a localized illumination is present (\emph{i.e.} a single Gaussian beam), Picoli's  theory of the photorefractive resonance\cite{pic89ol1362,Pic89ap3798} does not apply.
Hence there is a mere analogy between the phenomenon of inversion of the photorefractive response observed in Refs. \cite{Cha96ol1333,cha97apl2499}
and the photorefractive resonance \textit{stricto sensu}, as no background \ModifHerve{illumination} is present.
Especially, there is no particular reason why  the observed inversion intensity should  coincide with the resonance intensity $I_r$ measured by TWM and computed from the resonance theory.

\appendix{}
\section{Discussion of a previous theory}
Reference
\cite{uzd01ol1547} pretends to give a theory of the photorefractive resonance in the case of
a localized \ModifHerve{beam}. However, this theory is erroneous, and we intend to prove it in this appendix.

A mathematical error is found in the Appendix A of Ref. \cite{uzd01ol1547}.
The authors indeed state that `These equations  share the same solution (...); therefore they must have
the same coefficients', this argument is false.

Let us write a few elementary mathematics to justify this formally. Denote by $f$, $g$ and $h$
three arbitrary functions of $x$. Then set $\varphi={d\left(fg\right)}/{dx}$,
$\psi={d\left(fh\right)}/{dx}$.
Straightforward computation shows that
\begin{equation}
\frac{df}{dx}+\frac{dg/dx}g\; f =\frac \varphi g,\label{uz1}
\end{equation}
and
\begin{equation}
\frac{df}{dx}+\frac{dh/dx}h\; f =\frac \psi h.\label{uz2}
\end{equation}
The two equations (\ref{uz1},\ref{uz2}) share the solution $f$, hence
if the `property' involved in the cited paper were true,
we should have
\begin{equation}\frac{dg/dx}g=\frac{dh/dx}h,
\end{equation}
and hence $g$ and $h$ should be proportional, while they are arbitrary independent functions.
Consequently, the assumption $p(x)\propto n(x)$ in Ref. \cite{uzd01ol1547} is by no means justified.

Further, even if the property  $p(x)\propto n(x)$ is admitted,
a subsequent mistake is found in the reasoning.
Assume indeed that this statement is satisfied.   
Combining Eqs. (\ref{kuk4}-\ref{kuk6}),  looking for a stationary solution and
 neglecting diffusion, we obtain
\begin{equation}
\frac d{dx}\left(j_n+j_p\right)=0,\label{eqji}
\end{equation}
and consequently \begin{equation}\frac d{dx}\left(nE\right)=0,
\end{equation}
as in Ref. \cite{uzd01ol1547}.
Let us denote by  $n_\infty$, $p_\infty$ and $E_\infty$ the values of $n$, $p$ and $E$ at infinity in $x$, we get
$n={n_\infty E_\infty}/E$, and
$p={p_\infty E_\infty}/E$.
Reporting these  values into  system (\ref{kuk}), and taking into account
the fact that $j_n$ and $j_p$ do not depend on $x$ because their sum does not (Eq. (\ref{eqji}) ) and they are proportional due the assumption,
Eq. (\ref{kuk4}) and (\ref{kuk5})  yield respectively the two equations
\begin{equation}
\frac{p_t}{n_t}=\frac{e_n}{\gamma_nn}=\frac{e_nE}{\gamma_nn_\infty E_\infty},\label{prem}
\end{equation}
and
\begin{equation}
\frac{p_t}{n_t}=\frac{\gamma_pp}{e_p }=\frac{\gamma_p p_\infty E_\infty}{e_pE},\label{sec}
\end{equation}
which are not compatible, since  $E$
only depends on $x$.

In short: system (\ref{kuk}) does not admit any non-uniform solution with
$p(x)\propto n(x)$.
Consequently, Eq. (2) of Ref.  \cite{uzd01ol1547} is incorrect, and the consequences drawn from it are not founded.

\bibliographystyle{apsrev}
\bibliography{INP}
\end{document}